\documentclass[manuscript,screen]{acmart}

\usepackage{graphicx}
\usepackage{amsmath}
\usepackage{xcolor}
\usepackage{booktabs}
\usepackage{subcaption}
\usepackage{multirow}
\usepackage{tabularx}
\usepackage{array}
\usepackage{changepage}

\newcolumntype{Y}{>{\raggedright\arraybackslash}X}
\newcolumntype{P}[1]{>{\raggedright\arraybackslash}p{#1}}

\setcopyright{none}
\acmDOI{}

\title[Alignment Under Pressure]{Alignment Under Pressure: AR-HMD Support Tools for Action Teams}

\author{Jonathan Isaac Segal}
\orcid{0000-0002-8506-3785}
\affiliation{%
  \institution{Cornell University}
  \department{Information Science}
  \city{New York City}
  \state{NY}
  \country{USA}}
\email{jis62@cornell.edu}

\author{Jalynn Blu Nicoly}
\orcid{0000-0002-2897-5833}
\affiliation{%
  \institution{Colorado State University}
  \department{Computer Science Natural User Interaction Lab}
  \city{Fort Collins}
  \state{CO}
  \country{USA}}
\email{Jalynnn@colostate.edu}

\author{Huajie Cao}
\orcid{0009-0009-0347-3202}
\affiliation{%
  \institution{Michigan State University}
  \department{Media and Information}
  \city{East Lansing}
  \state{MI}
  \country{USA}}
\email{caohuaji@msu.edu}

\author{Soyon Kim}
\orcid{0000-0002-8232-9445}
\affiliation{%
  \institution{UC San Diego}
  \department{Computer Science and Engineering}
  \city{La Jolla}
  \state{CA}
  \country{USA}}
\email{sok020@ucsd.edu}

\author{Tauhid Tanjim}
\orcid{0000-0003-0491-5876}
\affiliation{%
  \institution{Cornell University}
  \department{Information Science}
  \city{New York City}
  \state{NY}
  \country{USA}}
\email{tt485@cornell.edu}

\author{Jonathan St. George}
\orcid{0000-0001-6810-4189}
\affiliation{%
  \institution{Weill Cornell}
  \department{Emergency Medicine}
  \city{New York City}
  \state{NY}
  \country{USA}}
\email{jos7007@med.cornell.edu}

\author{Kevin Ching}
\orcid{0000-0002-9008-5612}
\affiliation{%
  \institution{Weill Cornell Medicine}
  \department{Emergency Medicine}
  \city{New York City}
  \state{NY}
  \country{USA}}
\email{kec9012@med.cornell.edu}

\author{Francisco Raul Ortega}
\orcid{0000-0003-1285-6431}
\affiliation{%
  \institution{Colorado State University}
  \department{Computer Science}
  \city{Fort Collins}
  \state{CO}
  \country{USA}}
\email{F.Ortega@colostate.edu}

\author{Hee Rin Lee}
\orcid{0000-0002-5097-7309}
\affiliation{%
  \institution{Michigan State University}
  \department{Media \& Information}
  \city{East Lansing}
  \state{MI}
  \country{USA}}
\email{heerin@msu.edu}

\author{Angelique Taylor}
\orcid{0000-0003-1285-6431}
\affiliation{%
  \institution{Cornell Tech}
  \department{Information Science}
  \city{New York City}
  \state{NY}
  \country{USA}}
\email{amt298@cornell.edu}

\renewcommand{\shortauthors}{Segal et al.}

\begin{document}
\raggedbottom

\begin{abstract}
Team communication breakdowns represent a significant contributor to patient safety risks within action teams–\textit{defined as interdependent groups of specialized people who perform closely coordinated work under conditions of high workload, time pressure, and uncertainty}. 
Approximately 70\% of such instances lead to adverse patient outcomes due to the inherent complexities of tightly coordinated collaboration among healthcare workers operating under intense time pressure, uncertainty, and high cognitive load. 
While prior research has focused on maintaining shared cognition during these interactions, existing technologies largely prioritize individual task execution and decision-making, offering limited support for real-time team coordination. 
This study investigates the potential of augmented reality head-mounted displays (AR-HMDs) to address this gap by facilitating what we call ‘team alignment’ – the active maintenance of shared understanding regarding tasks, patient state, responsibilities, and ongoing clinical activity. 
Through an 11-month multi-phase qualitative study with ten healthcare professionals, we first elicited coordination challenges through semi-structured interviews complemented by real-time storyboard creation. 
Participants then engaged in reflection and refinement of these scenarios while contemplating the potential impact of AR-HMDs on their situation. 
Our findings revealed that breakdowns frequently arose when clinicians lacked sufficient contextual information, when assigned responsibilities did not align with available expertise, or when procedural progress was difficult to track – particularly during critical bedside activity. 
We subsequently developed the \textit{Team Alignment and Coordination Taxonomy} (TACT), encompassing information, expertise, procedural, and cognitive dimensions. 
By reframing coordination as this \textit{active maintenance} of alignment, our research shifts the design focus from individual decision support systems to holistic, team-level system interventions.
\end{abstract}
\maketitle



\section{Introduction}
\label{Introduction}

Effective coordination among multiple healthcare workers (HCWs) is fundamental to delivering safe and effective care in acute settings \cite{rosen_teamwork_2018, sarcevic_coordinating_2011, jeffcott_measuring_2008}. 
The emergency department (ED), in particular, presents a highly demanding environment where teams must rapidly diagnose, treat, and manage patients while navigating uncertainty \cite{leemeyer_decision_2020}, intense time pressure \cite{kusunoki_designing_2015}, and significant cognitive load \cite{grundgeiger_cognitive_2019}. 
These conditions contribute to HCW burnout, medical errors, and adverse patient outcomes \cite{hall_healthcare_2016, garcia_influence_2019, carver_medical_2026}.

Patient care is inherently team-based, involving HCWs with distinct roles who coordinate in real time to manage multiple patients simultaneously \cite{rosen_teamwork_2018,sarcevic_coordinating_2011}. 
Maintaining a shared understanding of patient status, task progress, and team priorities – a critical element of successful coordination – often proves difficult \cite{kusunoki_sketching_2015}.  
Traditionally, this shared understanding is established through continuous verbal communication as HCWs attend to diverse aspects of patient care \cite{sarcevic_coordinating_2011}.

These demands characterize acute care teams as \textit{action teams}--interdependent groups of specialized people who perform closely coordinated work to diagnose problems, plan and execute solutions, and reflect on opportunities for development. 
Furthermore, action teams operate under some combination of high workload, safety critical, time-critical, and uncertainty conditions \cite{edmondson_speaking_2003,kolbe_role_2011,johnson_team_2023,jamshad_human-robot_2026}.
Prior work has examined AR-HMDs within such settings; for example, Siebert et al., \cite{siebert_augmented_2026} evaluated role-specific AR-HMD decisions assistance for team leaders and medication nurses during simulated cardiac arrest, illustrating how technology can alter access to information and the coordination of responsibilities.
However, its standardized simulation and focus on two predefined roles lead to knowledge gaps on the effect of role-specific displays on coordination across changing teams, responsibilities, and clinical conditions.
For example, research on human-robot action teams shows that autonomous systems can reshape the distribution of responsibility and the practices through which teams respond to changing conditions.
Haripriyan et al., \cite{haripriyan_human-robot_2024} for example, found that teams adapted to a robot’s perceived limitations by developing workarounds and new roles .
This finding suggests that technologies introduced into action teams become part of their coordination structure and should align with established roles and practices.

Reliance on real-time verbal updates, “callouts,” and closed-loop communication highlights the importance of computer-supported cooperative work (CSCW) principles, specifically the need to align temporally distributed and partially observed information across collaborators \cite{reddy_finger_2002, cramton_mutual_2001}. 
However, this reliance is vulnerable to breakdowns, particularly during fast-paced procedures \cite{drews_interruptions_2019,mastrianni_alerts_2022}. 
Information can be delayed, missed, or misinterpreted, leading to misalignment in task execution and ultimately, compromised team understanding \cite{troyer_barriers_2020}. 
These disruptions aren’t isolated incidents; they represent recurring consequences of coordination in high-pressure team environments.

The challenges inherent in action teams coordination underscore the need for interactive systems that actively support the ongoing alignment of information across clinical roles. 
Existing technologies – such as checklists, cognitive aids, and alert systems – primarily focus on task execution and individual decision-making \cite{grundgeiger_cognitive_2019,mastrianni_transitioning_2023}. 
While valuable, these approaches often provide limited real-time assistance, team-level coordination, frequently rely on static or fragmented information, and demand that HCWs shift their attention across multiple devices during critical care \cite{pine_fragmentation_2012}.

Specifically, coordination challenges persist in fast-paced clinical environments where teams must keep patient status, task progress, and role responsibilities aligned across roles. 
This gap fuels the investigation into novel interactive systems that can facilitate continuous information alignment while preserving the essential hands-on interaction between HCWs and the patient, and maintaining direct sight of the clinical environment.  
Ubiquitous computing systems that embed information directly within the physical environment may facilitate this alignment by reducing the effort required to access, interpret, and share information across roles \cite{yarmand_enhancing_2024,warden_information_2024}. 
Among potential approaches, augmented reality head-mounted displays (AR-HMDs) represent a promising design space, though their actual effectiveness for improving team coordination remains largely unexplored.

Recent AR-HMD platform advancements make this design space increasingly important–combining see-through spatial displays with multimodal sensing, hand and voice interaction, shared spatial experiences, and AI models capable of interpreting aspects of the wearer’s visual and auditory context.
These capabilities could allow AR-HMDs to interpret and communicate evolving clinical activity in ways tailored to different team members.
At the same time, they raise important CSCW questions about how systems infer activity, distribute information across roles, maintain shared awareness, and preserve human authority.
Because our study did not explicitly examine AI-enabled AR-HMDs, we treat AI as one possible implementation mechanism rather than as a finding and focus on the coordination requirements that such systems would still need to satisfy.

This research investigates this unexplored territory. 
We conducted a multi-phase qualitative study with HCWs to understand how coordination challenges emerge during bedside care and how interactive systems might better facilitate shared cognition in these settings. 
Over 11 months, we conducted semi-structured interviews with physicians, nurses, and residents, to capture the nuances of team coordination in time-critical situations and identify the specific contexts where breakdowns occur.  
Building on these insights, we engaged participants in a co-design process using storyboards to explore how emerging technologies, including AR-HMDs, might facilitate coordination across roles.  
Rather than assuming AR-HMDs provide a solution, we used AR-HMDs as probes to elicit how HCWs reason about information sharing, role responsibilities, and real-time collaboration in high-pressure environments. 
Through this approach, we connected observed coordination challenges to design implications and examined the potential and limitations of AR-HMDs in action teams.

\begin{figure*}[!t]
    \centering
    \includegraphics[width=0.8\linewidth]{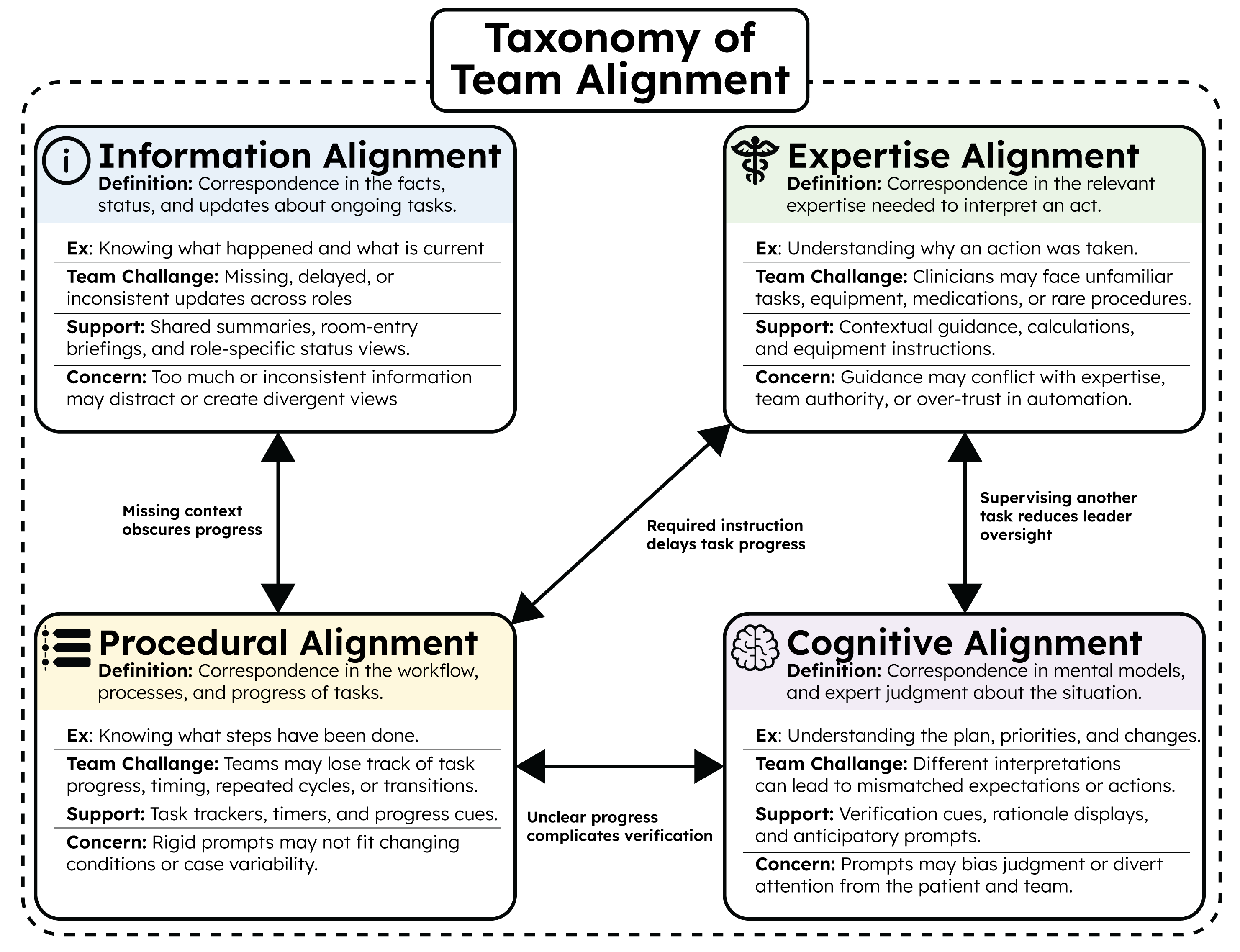}
    \caption{Overview of the Team Alignment and Coordination Taxonomy (TACT). Information, expertise, procedural, and cognitive alignment are analytically distinct but interrelated. Labeled arrows indicate breakdown pathways identified in participants’ accounts: missing context obscures progress, required instruction delays task progress, supervising another task reduces leader oversight, and unclear progress complicates verification.}
    \Description{Diagram of the Team Alignment and Coordination Taxonomy. Information, expertise, procedural, and cognitive alignment are arranged around team alignment. Labeled arrows show pathways from information to procedural alignment, expertise to procedural alignment, expertise to cognitive alignment, and procedural to cognitive alignment.}
    \label{fig:alignmenttaxonomy}
\end{figure*}

Figure~\ref{fig:alignmenttaxonomy} previews the four dimensions of TACT and the relationships among them identified through our analysis and described in the Findings.

This paper makes three contributions to CSCW and HCI research on team coordination:
\begin{itemize}
\item \textbf{TACT, a taxonomy of team alignment in action teams.} We introduce the Team Alignment and Coordination Taxonomy (TACT), which distinguishes coordination breakdowns across four dimensions of alignment: 1) information, 2) expertise, 3) procedural, and 4) cognitive alignment, and identifies relationships that technological interventions should preserve.
\item \textbf{Design Knowledge about AR-HMDs.} Through live co-design and storyboarding of AR-HMD design concepts, we provide insights into the potential and limitations of AR-HMDs for clinical team coordination, revealing key considerations for design, including risks related to situational awareness, communication, workflow integration, and clinical authority, emphasizing clinicians’ responsibility and discretion to make final care decisions.
\item \textbf{Design considerations for real-time team alignment across clinical roles.} We identify requirements for delivering evolving task information (e.g., patient state) and environmental changes (e.g., human actions, locations of supplies)  to the appropriate team members while preserving shared understanding across asymmetric roles and avoiding additional cognitive demands during time-critical care.
	\end{itemize}

\section{Background}
\label{Background}

\subsection{Shared Cognition in Healthcare Teams}
Emergency department (ED) workers, with interdependent roles, require tightly coupled, time-critical collaboration, where different actors maintain and act on distinct subsets of information \cite{bechky_gaffers_2006}.
Roles are typically assigned based on task assignment; thus, expertise (i.e., physicians, nurses, and residents) plays an important aspect of team effectiveness \cite{rosen_teamwork_2018,sarcevic_coordinating_2011}.
For example, Sarcevic et al., \cite{sarcevic_coordinating_2011} showed how trauma teams coordinate through differentiated roles and information exchanges as members enter, assess the situation, and contribute to ongoing care. 
Moreover, communication breakdowns contribute to preventable medical errors and adverse patient outcomes, making coordination critical \cite{carver_medical_2026}.
Prior research characterizes effective teamwork in acute care as teams adapting roles and synchronizing actions under time pressure \cite{jeffcott_measuring_2008}.

CSCW research frames coordination as the maintenance of shared cognition across distributed actors, who construct understanding through interaction with one another and with the artifacts that organize their work \cite{hutchins_cognition_1995,hollan_distributed_2000,kusunoki_sketching_2015}.
Kusunoki et al., \cite{kusunoki_sketching_2015} for instance, identified awareness of team members, tasks, elapsed time, and overall progress as important components of emergency medical teamwork. 
Because no single individual holds a complete view of tasks, closed-loop-communication requires teams to continually re-establish correspondence as situations evolve, as no individual has access to the full state of the situation at any given time \cite{kusunoki_sketching_2015,sarcevic_coordinating_2011,kusunoki_designing_2015}.
Team research has similarly used alignment to examine patterns in member attributes, shape coordination, and performance \cite{emich_team_2024,emich_better_2023}.
While this attribute-alignment research concerns team composition, other work has used \textit{team alignment} to describe the shared understanding and common ground that enable coordinated action \cite{avdiji_design_2018}.
Building on Avdiji et al., \cite{avdiji_design_2018}, who describe team alignment as the shared understanding and common ground that enable coordinated action, we extend this framing to dynamic clinical work by defining team alignment as the active maintenance of shared understanding regarding tasks, patient state, responsibilities, and ongoing clinical activity. 
This framing is consistent with research that treats teams as dynamic, multilevel, open systems that change over time \cite{emich_re-centering_2026} and may be beneficial to domains outside healthcare such as aviation \cite{mathieu_influence_2000}, construction \cite{bygballe_coordinating_2016}, and disaster response \cite{jones_designing_2020}.

Time pressure and changing patient conditions make team alignment difficult, as delays or mismatches in communication, interruptions, and shifting responsibilities can lead team members to form inconsistent interpretations of the situation \cite{drews_interruptions_2019, troyer_barriers_2020}.
Coordination breakdowns often occur not because information is unavailable, but because teams struggle to keep distributed and temporally evolving information aligned in real time across members with different responsibilities \cite{troyer_barriers_2020,reddy_finger_2002, cramton_mutual_2001}.
While prior CSCW research has described how teams build and maintain shared awareness, less is known about how interactive systems can enhance the real-time alignment of distributed information across team members in dynamic, multi-role settings without introducing additional complexity\cite{gonzales_visual_2016,gonzales_designing_2015,bardram_mobility_2005, reddy_finger_2002}.
Existing work therefore does not fully explain how ubiquitous systems can improve coordination in dynamic, multi-role environments where information remains partial, time-sensitive, and distributed across team members \cite{mastrianni_transitioning_2023, mastrianni_alerts_2022}.

\subsection{Existing Approaches to Clinical Action Teams}

Existing approaches attempt to enhance clinical work through tools such as checklists, cognitive aids, and alert-based mechanisms to enhance task execution and individual decision-making \cite{grundgeiger_cognitive_2019,gonzales_visual_2016,mastrianni_transitioning_2023}.  
However, these approaches remain limited in facilitating real-time, team-level coordination, as they often require proactive input, present information in static or fragmented forms, and require HCWs to shift attention across devices during time-critical tasks \cite{mastrianni_transitioning_2023}.
Gonzales et al., for example, found that existing electronic and paper-based documentation tools could fail during resuscitation, leading clinicians to record information on paper towels \cite{gonzales_visual_2016}
As a result, these approaches do not adequately assist with the continuous alignment of information across team members in fast-paced environments \cite{mastrianni_transitioning_2023,mastrianni_alerts_2022}.  
These limitations point to an opportunity for interactive systems that enhance real-time information alignment across roles without introducing additional complexity \cite{gonzales_designing_2015,mastrianni_transitioning_2023}.

\subsection{AR-HMDs as a Design Space for Coordination}
Augmented reality head-mounted displays (AR-HMDs) introduce one \textit{design space} for increasing coordination by enabling real-time, context-sensitive information to be embedded directly within the physical environment \cite{zhang_designing_2022}.
From this perspective, AR-HMDs have the potential to improve coordination by reducing the effort required to access and interpret information, allowing HCWs to align their understanding of the situation without shifting attention between disparate devices or representations \cite{haque_illuminating_2020,zhang_designing_2022}.
Recent AR-HMD platforms increasingly combine spatial displays with systems capable of perceiving and responding to the wearer’s immediate context.
Rather than functioning only as displays, these platforms may interpret evolving clinical activity and communicate relevant guidance across differentiated team roles.
However, the availability of these capabilities does not establish their reliability, appropriateness, or safety in clinical teamwork.

At the same time, AR-HMDs introduce challenges that complicate coordination.
Because these systems typically present information through individual, head-mounted displays, they can create asymmetries in displays, potentially disrupting shared awareness \cite{pelikan_operating_2018,randell_embedding_2016}.
These systems also risk increasing cognitive load and introducing visual clutter, particularly in high-pressure environments where attention is already constrained \cite{zhang_designing_2022}.
Thus, integrating AR-HMDs into existing clinical workflows remains difficult, raising concerns about usability, adoption, and the potential for new forms of coordination breakdown \cite{zhang_designing_2022, yarmand_enhancing_2024}.

Prior work has explored AR in healthcare for tasks such as training, visualization, and single-user guidance \cite{rampolla_augmented_2012,kyrarini_survey_2021}.
However, this work has largely focused on individual interaction or static use cases, rather than examining how AR-HMDs can improve coordination in dynamic, multi-user settings \cite{zhang_designing_2022,radu_survey_2021}.
As a result, it remains unclear how these systems affect the real-time alignment of distributed information across roles in action teams \cite{pelikan_operating_2018,bardram_mobility_2005}.
This gap motivates a critical examination of AR-HMDs as a design space for facilitating shared cognition, rather than assuming their effectiveness a-priori \cite{zhang_designing_2022,chung_negotiating_2023}.

\section{Methodology}
\label{Methodology}
We conducted an IRB-approved study (\#STUDY00008415) over 11 months using a multi-phase qualitative design to examine how coordination challenges arise in action teams and how HCWs reason about potential system support.
We designed the study to first elicit participants’ experiences with coordination breakdowns in practice and then use speculative AR-HMD scenarios and co-designed storyboards to explore how systems that embed information within the physical environment might enhance real-time coordination (see Figure \ref{fig:studyworkflow}).
This qualitative approach enabled participants to critically examine and reflect on possible configurations and consequences of AR-HMD use before a functional system was developed.
This structure allowed us to connect reported coordination breakdowns to design-relevant insights while surfacing both the opportunities and limitations participants associated with these speculative systems.
Throughout the interview, a researcher asked probing questions to better understand the nuances of AR-HMD use based on their lived experiences.

\subsection{Participants}
We recruited 7 male and 3 female (N=10)  healthcare professionals with expertise in emergency medicine through medical school mailing lists and professional referrals (see Table \ref{table:demo}).
Participant expertise ranged from physicians, to attending physicians, a director of clinical innovation, registered nurses, an assistant professor of emergency medicine, and a healthcare administrator with intensive care unit (ICU) nursing experience.
Participants reported prior experience across multiple institution types, including teaching (9), non-teaching (4), nonprofit (6), for-profit (3), urban (6), and rural settings (1).
Because many participants had worked in more than one setting, these categories are not mutually exclusive.
Participants reported low familiarity with AR-HMDs ($M = 2.4/5$ on a 5-point Likert Scale from 1 to 5) and received an \$18 Amazon gift card as compensation.

\begin{table*}[t]
\small
\centering
\caption{Participant demographics, including participant ID, age, gender, years of experience, specialty, and types of hospitals worked at: teaching (T), non-teaching (NT), for-profit (FP), non-profit (NP), urban (U), and rural (R).}
\label{table:demo}

\begin{tabular}{l r l r l l}
\hline
\textbf{Participant} & \textbf{Age} & \textbf{Gender} & \textbf{Experience} & \textbf{Specialty} & \textbf{Hospital Type} \\
\hline
P1  & 53 & M & 22 years  & Physician, Pediatric Emergency Medicine             & T, NP, U \\
P2  & 33 & M & 1.5 years & Healthcare Administrator, ICU Nurse                 & FP \\
P3  & 57 & M & 18 years  & MD, Emergency Medicine                              & T, NT, FP, NP, U, R \\
P4  & 51 & F & 21 years  & Attending, Emergency Medicine                       & T \\
P5  & 36 & F & 8 years   & MD, Pediatric Emergency Medicine                    & T, NT, NP, U \\
P6  & 37 & M & 8 years   & Director of Clinical Innovation, Emergency Medicine & T \\
P7  & 32 & F & 6 years   & Registered Nurse                                    & T, NT, FP, NP, U \\
P8  & 66 & M & 39 years  & MD, Emergency Medicine                              & T \\
P9  & 42 & M & 8 years   & Pediatric Emergency Medicine Attending Physician    & T, NT, NP, U \\
P10 & 38 & M & 13 years  & Assistant Professor of Emergency Medicine           & T, NP, U \\
\hline
\end{tabular}
\end{table*}

\subsection{Procedure}

\begin{figure*}[tbp]
    \centering
    \includegraphics[width=0.8\linewidth]{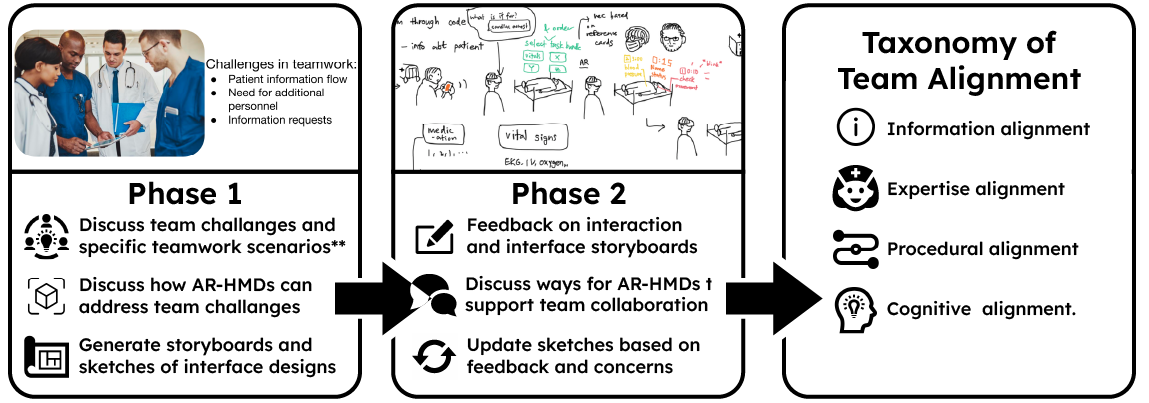}
    \caption{Overview of the two-phase study workflow. In Phase I, participants described coordination challenges and described scenarios from clinical practice. In Phase II, researchers and participants refined storyboards to explore how AR-HMDs might facilitate coordination. Thematic analysis of these data produced the Taxonomy of Team Alignment: information, expertise, procedural, and cognitive alignment.}
\caption*{\textbf{Scenarios:} (1) Walk through code, (2) Error detection, and (3) closing communication gaps.}
    \Description{Diagram showing a two-phase study followed by qualitative analysis. Phase I involves HCWs reflecting on coordination challenges. Phase II involves researchers and participants refining storyboards to explore how AR-HMDs might facilitate coordination. Analysis of the interviews and storyboards produces the Taxonomy of Team Alignment, comprising information, expertise, procedural, and cognitive alignment.}
    \label{fig:studyworkflow}
\end{figure*}

\begin{figure*}[tbp]
    \centering
    \includegraphics[width=\linewidth]{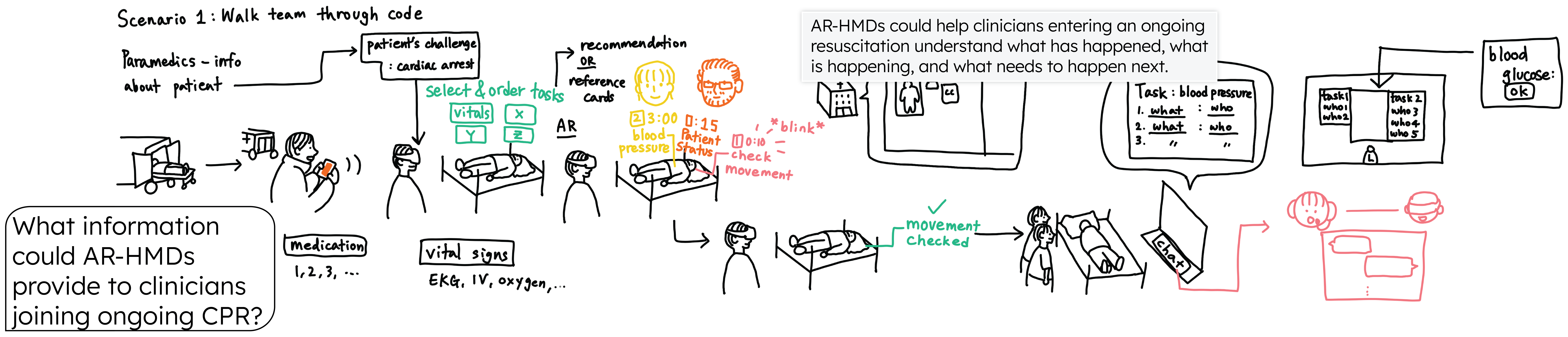}
    \caption{Example of the storyboard process used during the study. In Phase I, participants described coordination challenges and breakdowns from practice. In Phase II, the research team iteratively refined sketches with participants to explore how information might be shared, updated, and aligned across roles during action team procedures.}
    \Description{Screenshot of a collaborative storyboard session showing participants and researchers iteratively sketching interface ideas related to coordination during action team procedures.}
    \label{fig:storyboard}
\end{figure*}

We conducted one-hour semi-structured interviews with study participants over Zoom following the two-phase study procedure summarized in Figure~\ref{fig:studyworkflow}.
Before the interview, participants were asked to complete consent and demographic forms by email before the session.
At the beginning of the interview, the moderator confirmed participants' consent before recording began.
Each session included a moderator, a sketcher, and a note-taker.
The moderator guided the discussion, while the sketcher created real-time storyboards to externalize participants’ ideas (see Figure~\ref{fig:storyboard}).

We employed live, collaborative sketching to make participants’ assumptions about coordination and information sharing visible and to enable iterative refinement during the conversation.
Participants first responded to visual examples of possible technologies and scenario prompts concerning coordination challenges.
As they discussed these scenarios, the sketcher translated their accounts and speculative ideas into storyboards that participants could review and revise \cite{xue_co-constructing_2023,bowers_logic_2012}.
The resulting storyboards functioned as co-designed study artifacts that externalized participants' accounts and speculative ideas rather than finalized interfaces.
At the start of each session, we introduced the study goals and provided a brief overview of AR-HMD capabilities, including interaction modalities and example use cases, to ensure participants could meaningfully engage with the sketches (see Supplemental Materials).

\paragraph{Phase I: Understanding Coordination Challenges.} We began by showing participants brief video demonstrations and discussed the capabilities and limitations of AR-HMDs so that those with limited prior familiarity could meaningfully evaluate the scenarios.
Next, we  asked participants to describe coordination challenges from their own clinical practice, focusing on how information is communicated, updated, and aligned across team members during care tasks.
We used prior work on action teams coordination to sensitize the interview protocol, but did not assume these challenges were exhaustive \cite{sarcevic_coordinating_2011,gonzales_designing_2015}.
Instead, we prompted participants to describe situations where communication, role assignment, patient status updates, or procedural timing became difficult in their own experience.

\paragraph{Phase II: Reflecting on Storyboards with AR-HMDs.} During Phase I, the sketcher created the initial storyboards in real time as participants described coordination challenges from clinical practice.
In Phase II,  participants revisited these storyboards and considered how ubiquitous systems such as AR-HMDs might alter the situations they had described.
Rather than assuming that AR-HMDs provided a solution, we used the storyboards to examine how participants reasoned about information sharing, role responsibilities, and real-time collaboration.

We iteratively refined the storyboards during this discussion so that they accurately reflected participants’ reasoning and made design tradeoffs visible \cite{taylor_towards_2024}.
Example prompts included: “How should information be displayed for different team members?”, “What information should be shared across the team?”, “How should time-sensitive information be communicated?”, and “How should displays differ across roles?” (see Supplemental Materials for the full protocol).
This process allowed us to examine both potential benefits and limitations of AR-HMDs in improving coordination.

We selected this qualitative, multi-phase approach because direct observation of coordination in action teams is often constrained by access, privacy, and the unpredictability of time-critical events. 
Semi-structured interviews combined with the sketches allowed us to capture both participants’ lived experiences and their reasoning about coordination assistance in context. 
This approach has been shown to be effective for eliciting tacit knowledge and uncovering coordination mechanisms that may not be directly observable in real-time settings \cite{xue_co-constructing_2023}.

\subsection{Analysis}
We recorded and transcribed all interviews using Microsoft Stream and analyzed the data using an iterative thematic analysis approach \cite{braun_thematic_2012}.
This involved two researchers independently reviewing the transcripts and coding participants’ accounts of coordination challenges alongside their responses to the AR-HMD scenarios and storyboards.
The analysis followed information across roles, examined its interpretation, and documented shifts in shared understanding over time.
Through iterative comparison and discussion, the researchers reached agreement on a final coding structure centered on situational knowledge, expertise and responsibility, procedural progress, and the verification of consequential information.
Rather than organizing the findings around predefined technological features, we examined how participants’ experiences and design ideas revealed underlying coordination mechanisms and where AR-HMDs might enable or disrupt them.
We continued this process until no new themes emerged in the later transcripts.

\section{Findings}

Our Taxonomy of Team Alignment captures four types of alignment categories that characterize challenges HCWs faced and how AR-HMDs might mitigate those challenges: (1) Information, (2) Expertise, (3) Procedural, and (4) Cognitive  alignment.
Unlike prior taxonomies related to clinician teams, ours captures unique challenges emergency medicine teams face and future opportunities to leverage AR-HMDs as hands-free devices to improve team cognition.
Our participants framed AR-HMDs as most useful for addressing this taxonomy to mitigate conflicting knowledge, actions, and decision-making beyond.
This framing emphasizes contextualizing understanding of the environment, patient and HCW states, as well as transitions over time through knowledge recording and sharing to balance interface visibility and interference.

\subsection{Information Alignment: Bridging Knowledge Gaps in Pre-Entry Situation Briefing} 
\textit{Information alignment} refers to accurate numerical and historical information about patients as well as the consistency of retained knowledge among the team to take appropriate actions at the right time, as described by four study participants.
This challenge is especially important upon arrival to patient rooms, when HCWs have no information about the patient, their condition, or actions already taken by the clinical team.
Particularly in crowded, multi-service emergencies, where incoming HCWs needed to understand patient status, team structure, and current priorities before they could act effectively, briefing teams before arrival could serve as a vehicle to enable rapid onboarding of HCWs iteratively joining the team.
To address the problem, participants imagined AR-HMDs as a \textit{information access layer} to bridge this gap, not only as bedside displays, but as transitional coordination tools that could prepare incoming HCWs before they physically entered the room.
For example, P9 described a use-case of AR-HMD visual displays based on their experience in a pediatric obstetrics, neonatal intensive care unit of an emergency department: 
\begin{quote}
\itshape
\textit{``The augmented reality could be interesting from a team perspective because if the information can be displayed on some of these headsets [\ldots] prior to the teams arriving, they may be able to have information ahead of time. I talked about how the [obstetrics and gynecology] and the [neonatal intensive care unit] team have to come from across the street. Well, if they have these headsets and as soon as they get the call, they're able to put on the headset and receive information in real time from the team that's actually there, there may be some utility in terms of creating that mental model and making sure that everyone's on the same page. So that way, as soon as they get to the room, they know exactly what's going on. And that may be a live feed of what's going on in the room.''}
\end{quote}

This account also shows how \textit{missing context obscures progress}.
Clinicians arriving without an account of what had already occurred could not immediately interpret the current state of care or determine how their responsibilities related to the ongoing procedure.
As a result, they required additional briefing, repeated questions or assessments, and risked delaying, duplicating, or overlooking time-sensitive actions.
\subsection{Expertise Alignment: Augmenting Role-Specific Guidance}
\begin{figure*}[t]
    \centering
    \begin{subfigure}[b]{0.42\textwidth}
        \centering
        \includegraphics[width=\linewidth]{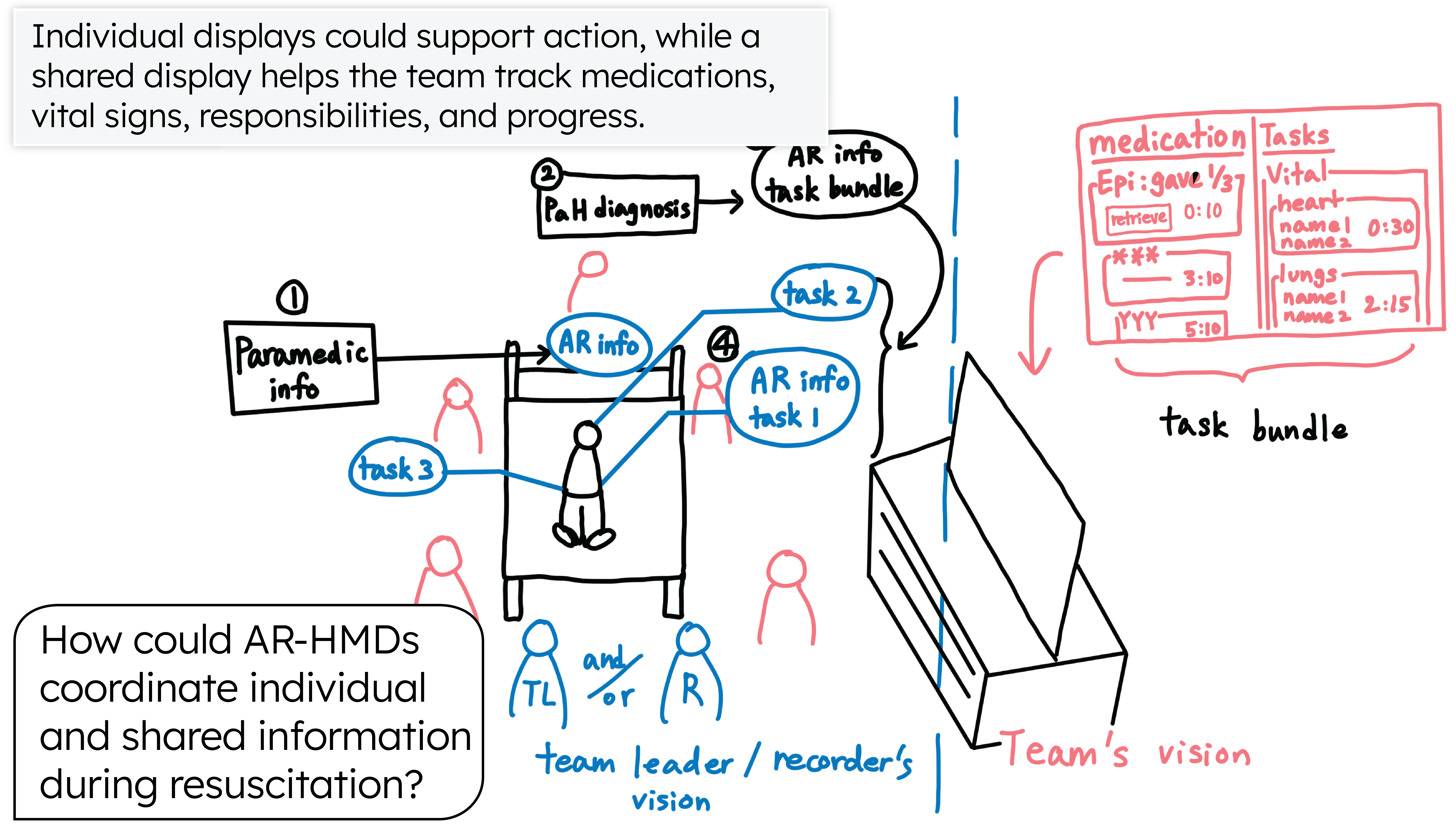}
        \caption{Storyboard exploring timely updates during CPR and resuscitation.}
        \Description{Hand-drawn storyboard showing CPR timing, medication intervals, and task tracking during resuscitation.}
        \label{fig:sb1}
    \end{subfigure}
    \hfill
    \begin{subfigure}[b]{0.52\textwidth}
        \centering
        \includegraphics[width=\linewidth]{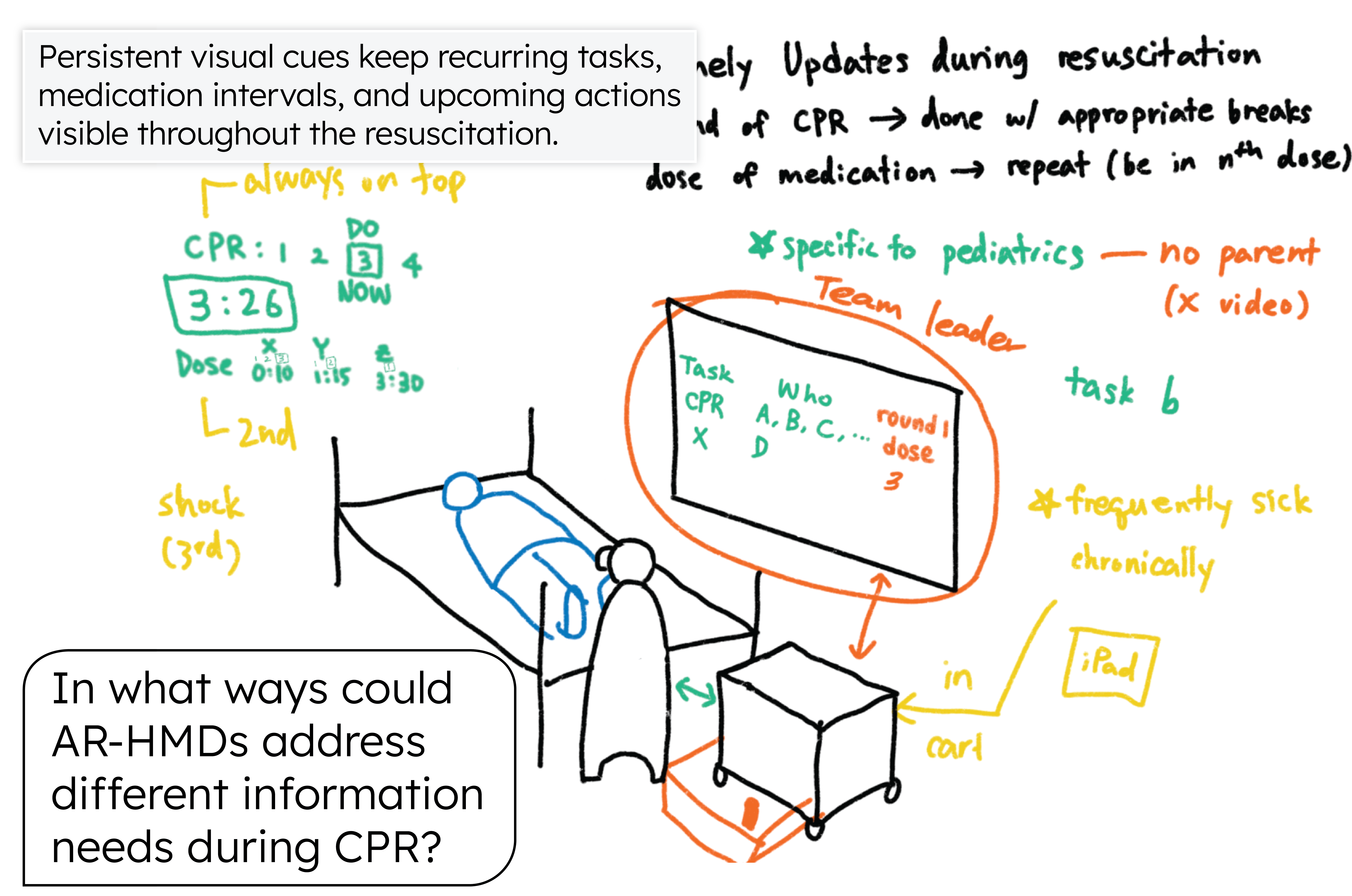}
        \caption{Storyboard exploring role-specific task bundles and shared versus individualized views.}
        \Description{Storyboard showing different team members accessing task bundles and patient information, illustrating shared and role-specific views.}
        \label{fig:sb2}
    \end{subfigure}
    \caption{Example storyboards showing how participants imagined how AR-HMDs could augment the alignment of assigned tasks to appropriate members of the care team.}
    \Description{Two storyboards illustrating coordination scenarios, including task timing, patient monitoring, and distribution of information across team roles.}
    \label{fig:exp_alignment}
\end{figure*}

\textit{Expertise alignment} arises as a nuanced phenomena when HCWs are aligned with tasks that do not align with their training (e.g., a pharmacist assigned to diagnose patients) or other factors impede their task progression, as highlighted by four participants (see Figure \ref{fig:exp_alignment}).
This issue often arises due to lack of familiarity of those joining the team and there is limited time for introduction in \textit{action teams}.
Instead, HCWs spring into action when assigned a task, sometimes without the team leader understanding that individual’s expertise and background.
Additionally, this taxonomy captures when HCWs are aligned to tasks within their trained skillset, sometimes they experience deskilling due to lack of exposure to patients with similar conditions.
Furthermore, unevenly distributed roles (i.e., tasks), particularly in large teams, can impede team coordination, causing the team leader to balance many responsibilities, including leading, taking action, supervising, and teaching simultaneously, causing cascading mismanagement of the entire team.
Participants noted that, although these situations occur rarely, patients have severe conditions which require HCWs to be prepared at any time to treat rare conditions; otherwise, patient safety risks arise.
P8 reflected on a pediatric emergency patient case they experienced that reinforces this taxonomy, in a rare procedure in a low-resourced facility:
\begin{quote}
\itshape
\textit{``So sometimes [\ldots] in [a low-resourced emergency department], it's particularly challenging as the [HCW with pediatric emergency care experience] because we might be the only person in the room with significant experience caring for a critically ill child. And so you have to be [\ldots] Jack of all trades, playing several roles at once while also on the fly explaining [to] other people how to play their role in the context. So it can be pretty overwhelming. Thankfully, a lot of problems with pediatric emergency medicine are [\ldots] low frequency but high severity, high morbidity events. And so it's often not the case that you encounter these scenarios, but every time you do encounter the scenarios because of their relatively low frequency compared to adult counterparts [\ldots].''}
\end{quote}

To address the aforementioned concerns, participants emphasized the benefits of using AR-HMDs for real-time training through customized displays based on HCW roles based on equipment use, medication dosage calculation and administration, and task reminders for iterations tasks in particular.
By translating expert guidance into prioritized, actionable, role-specific notifications, AR-HMDs can bridge knowledge gaps between experience and in-experiences HCWs to maintain common ground, alignment of evidence-based practices, and interpretation of best practices to act upon it using sound clinical judgment.
The goal of taxonomy is not intended to replace expert HCW guidance, but to increase accessibility of evidence-based knowledge while preserving clinical authority, meaning clinicians’ responsibility for directing patient care, and legal compliance, which is particularly useful when managing multiple patients.
P2’s account of managing two patients with junior clinicians to treat a patient in need of sedation to treat a dislocated joint and a patient with a severe eye injury:
\begin{quote}
\itshape
\textit{``We were left having to [\ldots] balance these 2 patients. One of them was [\ldots] someone that was requiring moderate sedation for a joint dislocation. And at the same time [we] had a patient that required an emergent procedure to save their eye, after falling and hitting their head. [It was] made more complicated by the fact that the patient was intoxicated, belligerent, violent, [\ldots] and with significant psychiatric comorbidities. And so [\ldots] that's an example of having [to] really balance what our priorities are. And at that point the level of experience of my team definitely made a difference.''}
\end{quote}

\subsection{Procedural Alignment: Enhancing Shared Mental Models of Clinical Tasks}
\begin{figure*}[t]
    \centering
    \begin{subfigure}[b]{0.42\textwidth}
        \centering
        \includegraphics[width=\linewidth]{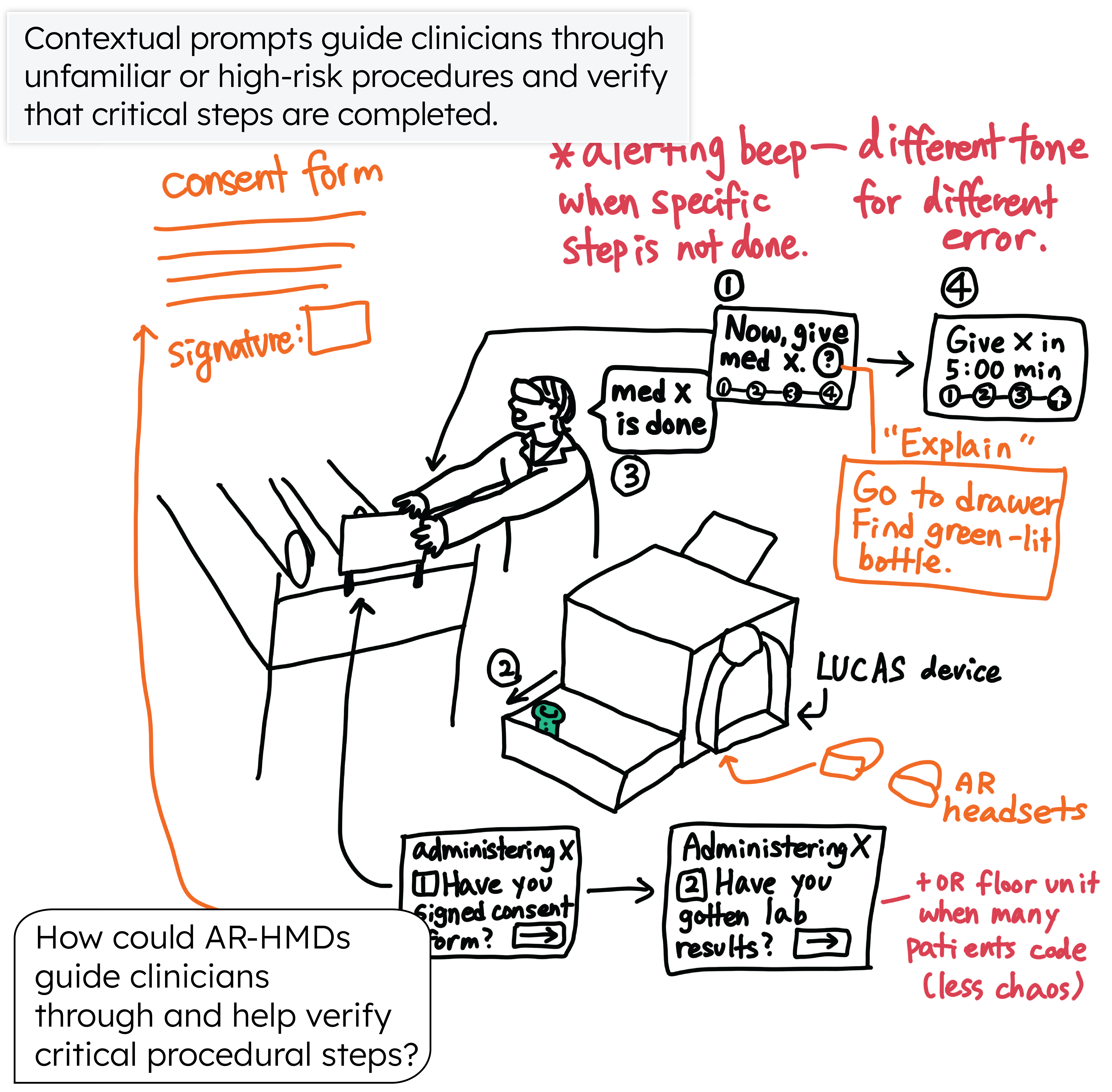}
        \caption{Storyboard exploring guidance for rare procedures.}
        \Description{Storyboard showing step-by-step guidance for a rare medical procedure, including medication timing and equipment retrieval.}
        \label{fig:sb3}
    \end{subfigure}
    \hfill
    \begin{subfigure}[b]{0.52\textwidth}
        \centering
        \includegraphics[width=\linewidth]{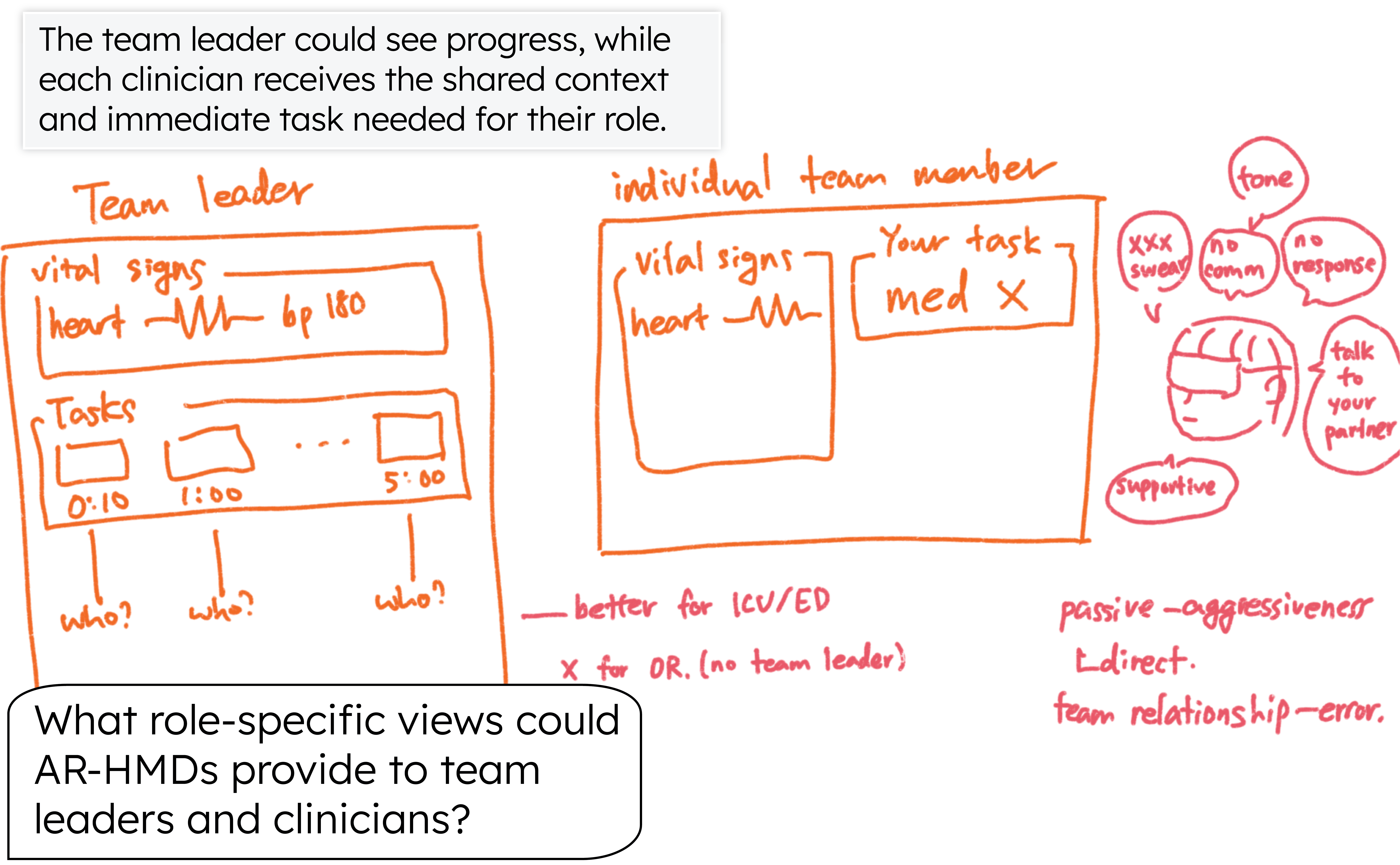}
        \caption{Storyboard contrasting team-leader and individual views.}
        \Description{Storyboard comparing a team leader’s overview display with an individual HCW’s task-specific display.}
        \label{fig:sb4}
    \end{subfigure}
    \caption{Storyboards illustrating how AR-HMDs could support procedural alignment by making task progress, role-specific responsibilities, and critical procedural steps visible across team members while preserving differentiated views for team leaders and clinicians performing focused tasks.}    \Description{Two storyboards highlighting differences in how information is presented across roles, including centralized and individualized views.}
    \label{fig:proc_alignment}
\end{figure*}

\textit{Procedural alignment} is defined as the shared mental model of task status, progression, and conflicts being on the same page when it comes to the steps of the algorithm, as indicated by six participants (see Figure \ref{fig:proc_alignment}).
Conflicts refer to issues faced by the team during active care tasks and misalignment of decision-making stemming from retained clinical knowledge.
During cardiopulmonary resuscitation for example, clinical teams must track medication doses, compression cycles, rhythm checks, task completion, and whether key updates have been communicated.
When verbal updates are unclear or missed by the team, the shared mental model of the team breaks down, including coordinating their actions with others effectively and quickly.

Procedural misalignment could also result from expertise misalignment.
This situation shows how required instruction delays task progress and how supervising another task reduces leader oversight. When the resident required direct assistance with an assigned task, the team leader had to leave their coordinating position and could no longer monitor the broader resuscitation.
P1 described pausing the primary survey and leaving the foot of the bed to help a resident immobilize the cervical spine:

\begin{quote}
\itshape
\textit{``We had not yet finished the primary survey because I had sort of had to pause to help the resident immobilize the cervical spine. [\ldots] I needed to leave my position from the foot of the bed as the team leader. And the foot of the bed is the ideal location for the team leader because it gives you a good vantage point for the entire resuscitation--the environment, your patient, the team, [and] the vital-sign monitor. [\ldots] So I had to leave that vantage point in order to go to the head of the bed to assist.''}
\end{quote}

This account shows that the consequence extended beyond the resident’s need for guidance.
Providing that guidance interrupted procedural progress and temporarily reduced the leader’s overview of the patient, team, and clinical environment.
\begin{quote}
\itshape
\textit{``I do think that having speech recognition in real time, [and] monitoring closed-loop communication, I think that it could be a really positive reinforcement [loop] [\ldots]. If you know that when you placed an 18 gauge IV and you're supposed to say `I placed an 18 gauge,' sometimes people don't do it, but if doing it is going to then pop up on the screen, [and] a green light pops up on the arm where the 18 gauge IV was placed so that everybody knows the task was done, it may also be that it reinforces people's closed-loop communication.''}
\end{quote}

Further enhancing procedural alignment, AR-HMDs could translate spoken updates involving time-sensitive numerical information into visual feedback that helps team members confirm that the information was communicated and understood accurately. 
Because clinical teams rely on actively listening and speaking to others throughout the code, it is important for AR-HMDs to have a contextual understanding of when it is or is not appropriate to communicate using speech.
Instead AR-HMDs can leverage visual feedback to update displays as an ambient listening device from sensing the environment and headset wearer, thereby confirming task progression automatically.
Our findings suggest AR-HMDs could enhance procedural alignment temporally through visual displays of HCW roles to toggle shared state information \textit{globally} for the team leader and \textit{locally} for everyone else.
Through making spoken updates visible without overloading every team member, this approach reinforces closed-loop communication while reducing the need to repeatedly reconstruct the procedural timeline.
P3 also argued that procedural information can serve as a shared mental model across roles rather than shown uniformly to everyone: 
\begin{quote}
\itshape
\textit{``The screen that the entire team can see should be very simple, because the entire team is overwhelmed and they're also focusing on their own one task, so that screen needs to be super simple. It needs to just have [\ldots] a quick number so that we know where we're at. The screen that the team leader can see should have more detail about, for example, on the medication line, these are the meds that have been given at these times, but for the team, it's just like this is the medication being given now.''}
\end{quote}
\subsection{Cognitive Alignment: Enhancing Clinical Judgement Through Verification}

Participants shared the importance of verification of time-sensitive information to reduce cognitive load of establishing common ground, which we refer to as \textit{cognitive alignment}.
For example, HCW may need to verify patient status, algorithmic steps; medication type, doses, and timing.
This taxonomy is essential because HCWs need to check critical information without leaving the bedside or interrupting hands-on care; otherwise, care delays persist.
To address this problem, four participants favored a {cognitive layer} with low-latency, high-confidence based on evidence-based practices from AR-HMDs, such as notifications about when to administer medication, dosage confirmation, and patient status updates at the right time.
These cases also show how \textit{unclear progress complicates verification}, making it harder for clinicians to confirm what has been completed, what remains pending, and whether the team's understanding of the situation remains aligned.
P4, in particular, described the impact AR-HMDs can make during medication dosing errors to explain why real-time verification matters:
\begin{quote}
\itshape
 \textit{``So human error happens pretty frequently. I myself as an intern once ordered the wrong dose of Ativan on a patient who was actively seizing. Ativan is an anti-seizure medication and thankfully, [\ldots] I don't remember if it was the nurse or the senior resident, someone caught it, but I'm actually kind of shocked now in retrospect that the medical record, like the ordering system, didn't catch it. And I think that is something that [a system] could definitely help in. Dosing is always a problem and then [\ldots] the timely updates [are] exactly what I think would be most helpful during an active resuscitation.''}
\end{quote}

Participants provided careful reflection of what AR-HMDs should and should not be capable of– specifically, they emphasized AR-HMDs as a verification tool, not as an autonomous decision-maker.
More specifically, AR-HMD decisions should be overridden when its recommendation does not align with a HCW’s knowledge, specifically when double-checking clinical knowledge is time-sensitive due to severe patient state.
Instead, HCW should maintain autonomy and authority and AR-HMDs can assist in differential dosages against the current procedural state, medication timing, and patient status, without shifting interpretive authority away from clinical leaders.
P7 emphasized this boundary when describing why the team leader may need the most complete view of the care team through a headset:
\begin{quote}
\itshape
\textit{``Well, because that's [\ldots] you're making all the decisions. You know, that's your job, is to analyze the situation and direct the team appropriately. I don't see how it could be any other way. I mean, [because] you're the decision maker. I think you should have all the information displayed to you. And then you can make the decisions.''}
\end{quote}

\subsection{Concerns about AR-HMDs in Challenging Environments}
Although participants acknowledged the benefits of using AR-HMDs during emergency care procedures, they also reflected on concerns based on the factors that characterize these environments as dynamic and chaotic.
Specifically, seven participants expressed concerns about maintaining situational awareness in terms of attending to and hearing others in loud fast-paced settings, such as the emergency department.
For example, P8 describes a situation in which they participated in a research study that used Google glasses in a code situation to express how it interfered with team coordination, conflicting with the study goal:
\begin{quote}
\itshape
\textit{``I have actually been part of a study back in fellowship where we were thinking of using [\ldots] Google Glass for the code situation and wear it. For me, it was a little disorienting to try to have that on my face while I'm also trying to keep an eye on things. And then also [\ldots] I need to be able to hear [\ldots] information from the team leader. Essentially [\ldots] I needed to have a better awareness of everything that is going on around me and have all of my hearing and my ability to communicate with the rest of my teammates.''}
\end{quote}

Interactions with AR-HMDs are inherently multimodal using speech, gaze, and gesture to update the system state.
As a result, HCWs must maintain their sense under the guise of AR-HMD interactions, including visibility of the room, listening to team conversations, speaking without interruptions, and tracking the team state over time.
Therefore, AR-HMDs need to provide sparse, glanceable, and dismissible information without occluding the environment or competing with verbal communication.
This finding reveals a central design constraint–that is, AR-HMDs need to function reliably to sense human verbal and nonverbal cues in noisy, crowded, and rapidly changing environments.
Thus, AR-HMDs need to provide sparse, glanceable, and dismissible notifications while avoiding occluding the user’s environment and conflicting or missing their verbal cues.
P3 emphasized that AR-HMDs should provide immediately relevant information without distracting clinicians from their situational awareness:

\begin{quote}
\itshape
 \textit{``I think AR presents a number of interesting potential possibilities, [including] the ability to display data in a way that gives me useful information but also doesn't distract me from my situational awareness. [\ldots] Having a clock running [could] give me a sense of how well we're doing on tasks. [\ldots] If we're supposed to give epinephrine every three minutes [or] a pulse check has to happen every so often, then I can know when that pulse check is supposed to happen or if we're overdue. [\ldots] Rather than having to either make someone leave the bedside to go to a desktop computer to gather information and then come back and tell me what the relevant information is, [the headset could provide] a code-relevant short list of good data visualization that pops up and tells me this patient's DNR status, what their allergies might be, and what their most relevant past medical history might be.''}
\end{quote}

The four forms of alignment introduced here are distinct, but not isolated.
Across participants' accounts, breakdowns often propagated across dimensions: missing context obscures progress, required instruction delays task progress, supervising another task reduces leader oversight, and unclear progress complicates verification.
Interventions intended to improve one form of alignment could also threaten another by diverting attention or interfering with communication and awareness of the team.
Figure~\ref{fig:alignmenttaxonomy} summarizes these pathways across the four dimensions of team alignment.

\section{Discussion}
\subsection{Design Considerations for AR-HMD Enhanced Team Alignment}
This study contributes a taxonomy of team alignment that explains how clinical teams maintain shared understanding by preserving sufficient correspondence among role-specific and time-sensitive forms of knowledge as care unfolds.
The taxonomy distinguishes information, expertise, procedural, and cognitive alignment as distinct but potentially interrelated dimensions through which teams maintain a coherent understanding of care.
By connecting these four dimensions of alignment in system design, the taxonomy provides a basis for examining how technologies affect coordination across clinical roles.
These dimensions identify corresponding opportunities for assistance: AR-HMDs may facilitate information alignment through concise pre-arrival or room-entry summaries, expertise alignment through guidance for unfamiliar equipment or rare procedures, procedural alignment through displays of task progress and upcoming transitions, and cognitive alignment through bedside verification of consequential information.
The contribution is not the general observation that different clinical roles may require different interfaces; rather, it is identifying how differentiated views must preserve alignment as information, available expertise, procedural progress, and clinical interpretation changes.

The taxonomy also informs research beyond acute care, including human-machine teaming, in which technologies increasingly participate in dynamic, multilevel team systems \cite{emich_re-centering_2026}.
Such arrangements require coordination across differences in teams knowledge and authority, and interpretability of AR-HMDs.
The taxonomy can help researchers examine whether an AR-HMD intervention preserves alignment across information, responsibilities, progress, and clinical judgment.Rather than maximizing the information visible to every team member, ubiquitous systems should maintain correspondence among differentiated views.
The taxonomy therefore offers a relational account of the \textit{active maintenance of team alignment}, showing the trajectory of breakdowns in one form of alignment and propagation of information into others.

Prior work establishes that team cognition is distributed across team members and the artifacts through which they organize activity \cite{hutchins_cognition_1995,hollan_distributed_2000}.
Research on common ground similarly explains how collaborators make decisions from incomplete information \cite{cramton_mutual_2001}, while work on role-based coordination shows how responsibility shapes access to information and hierarchy \cite{bechky_gaffers_2006}. 
These perspectives inform our view that coordination depends not on identical knowledge, but on maintaining sufficient correspondence among team members’ individual understanding.
However, they do not distinguish the forms of alignment that teams must maintain as conditions, responsibilities, and team composition change.

Unlike prior research on team alignment that captures team member characteristics, our taxonomy characterizes alignment as a continual process involving patient information, available expertise, procedural progression, and clinical interpretation.
For example, prior research described alignment of member attributes to capture team composition, demonstrating the influence of coordination on performance \cite{emich_team_2024,emich_better_2023}.
In contrast, our taxonomy distinguishes the forms of alignment that teams must maintain and identifies specific pathways through which a breakdown in one dimension may affect another.
For example, a clinician who joins after treatment has begun may be unable to interpret procedural progress without knowing what has already occurred, while a clinician assigned an unfamiliar task may require guidance, and clinicians making consequential decisions may require access to verification before acting.
Alignment must therefore be reconstructed as the situation and the team's responsibilities change.
This view corresponds with research that characterizes teams as dynamic, multilevel, open systems whose organization and interactions develop over time \cite{emich_re-centering_2026}.
While the aforementioned work explains how teams evolve, our taxonomy specifies the forms of alignment that must be reconstructed as conditions, responsibilities, and team composition change, while showing how breakdowns can propagate across them.

Prior work often treats alignment as convergence in team members’ mental models or the development of shared understanding and common ground \cite{mathieu_influence_2000,avdiji_design_2018}.
TACT extends this work by showing that alignment does not require every team member to receive identical information.
Action teams are organized through differentiated roles and authority \cite{bechky_gaffers_2006}, and team members therefore require different levels and forms of information.
A team leader may need a detailed account of how care has progressed, while another clinician may need only enough context to perform an immediate task.
These differences are not inherently coordination failures.
We therefore conceptualize misalignment as occurring when distinct views no longer correspond sufficiently with one another to assist coordinated action.

The taxonomy builds on prior work on distributed cognition \cite{hutchins_cognition_1995}, common ground \cite{cramton_mutual_2001}, and differentiated roles \cite{bechky_gaffers_2006} by specifying the forms of correspondence that must be maintained across team members’ distinct perspectives, conceptualizing alignment as an ongoing process involving information, expertise, procedural progress, and expert judgement.
We describe this design objective as coordinated differentiation–role-specific views may vary in detail but should preserve enough shared structure to maintain team understanding and orient them toward a common course of action.
These findings suggest five related design considerations for systems intended to support team alignment: 1) facilitating rapid onboarding for clinicians joining care in progress; 2) customizing guidance beyond formal roles; 3) making temporal dependencies and procedural progress visible; 4) maintaining clinical authority and interpretability; and 5) preserving shared reference points across differentiated views. 
Table~\ref{tab:design_implications} summarizes the coordination need associated with each consideration and its corresponding implication for system design.
Rapid onboarding requires concise context about prior activity, current patient state, and team responsibilities, while supporting procedural alignment requires making completed actions, incomplete work, medication timing, and upcoming transitions visible as care progresses.
Such guidance should account for the possibility that  assigned roles do not fully represent a clinician's expertise or the demands of the case.
Cognitive alignment requires enabling verification without diverting attention from bedside care or displacing professional judgment.
Because clinical roles carry different decision-making responsibilities, systems should not treat all users as interchangeable decision-makers or convert recommendations into directives.
Ubiquitous systems that prevent clinicians from questioning or overriding outputs, or that shift interpretive authority away from accountable team members, may create cognitive misalignment even when the information presented is accurate.

TACT may also support comparison of how technological interventions could facilitate or disrupt team alignment.
Prior work shows that electronic and paper-based documentation tools can fail during resuscitation, disrupting clinicians’ ability to track in procedural progress and requiring them  to improvise alternative recording practices \cite{gonzales_visual_2016}.
Prior research also suggests that individually presented AR-HMD information may leave team members uncertain about what others can see \cite{siebert_augmented_2026}.
TACT therefore shows that role-specific views must preserve shared reference points to patient status, procedural progress, and team priorities, while interfaces should avoid interfering with speech, hearing, movement, visibility, and awareness of others.
The value of an intervention lies not in maximizing the information presented, but in providing sufficient role-relevant information needed to sustain alignment without adding unnecessary cognitive demands.
Future work should examine whether functioning systems can maintain correspondence between their computational representations of activity and clinicians’ understandings as conditions evolve.

\begin{table*}[t]
\small
\centering
\caption{Design considerations for AR-HMD enhanced team alignment in action teams.}
\Description{A three-column table connecting the forms of team alignment to coordination needs and design implications.}
\label{tab:design_implications}

\renewcommand{\arraystretch}{1.2}
\setlength{\tabcolsep}{6pt}

\begin{tabularx}{\textwidth}{
>{\raggedright\arraybackslash}p{0.16\textwidth}
>{\raggedright\arraybackslash}p{0.28\textwidth}
>{\raggedright\arraybackslash}X
}
\toprule
\textbf{Design consideration type} &
\textbf{Coordination need} &
\textbf{Design implication} \tabularnewline
\midrule

Rapid onboarding
& Clinicians joining care in progress need to understand patient status, prior interventions, current priorities, and team organization.

& Systems should provide concise, time-stamped summaries of prior activity, current patient state, incomplete work, and team responsibilities before or as clinicians enter ongoing care. \tabularnewline

Customization beyond formal roles
& Responsibilities may not correspond with clinicians' training, recent experience, or familiarity with rare procedures.
& Systems should provide adjustable guidance based on assigned responsibility, recent experience, and task familiarity without assuming that formal roles fully represent expertise. \tabularnewline

Mitigating temporal propagation
& Teams must maintain a shared account of completed actions, current tasks, upcoming transitions, repeated cycles, and incomplete work.

& Temporal dependencies and procedural progress should be made visible through real-time capture of task completion, medication timing, incomplete work, and changes in the current plan. \tabularnewline

Maintaining authority and interpretability
& Clinicians must verify consequential information while remaining engaged in bedside care and retaining authority over decisions.

& Rapid verification of medication, patient, and procedural information should preserve source and uncertainty while allowing clinicians to question, reject, or override system outputs. \tabularnewline

Preserving shared reference points
& Different roles require different levels of detail while still needing sufficient common ground to interpret one another's actions.
& Differentiated views should preserve shared reference points across role-specific interfaces rather than present identical information to every team member. \tabularnewline

\bottomrule
\end{tabularx}
\end{table*}

\subsection{Opportunities and Constraints of AR-HMDs for Team Alignment}
Our third contribution identifies the opportunities and constraints of AR-HMDs as one possible mechanism to improve team alignment, in comparison to alternative solutions.
Instead of viewing AR-HMDs as a comprehensive solution to action team coordination, participants identified circumstances in which head-worn displays could improve access to role-relevant and time-sensitive information while allowing clinicians to remain engaged with the patient and clinical environment.
Across these forms of alignment, the opportunities arise from making contextual information available within the clinician's field of view rather than requiring attention to shift toward a separate device.
However, attention, hearing, movement, communication, and visibility are not peripheral usability concerns, but resources through which action teams coordinate.
The taxonomy can therefore help identify when an intervention intended to improve information access may create new breakdowns by interfering with speech, hearing, visibility, attention, or awareness of other team members.
An AR-HMD that disrupts these resources may create information or procedural misalignment even when the information it presents is accurate.

Recent AR-HMD developments, such as Snap Specs and Meta Ray-Ban Displays, make these opportunities increasingly feasible while reinforcing the need for team-level design guidance.
By sensing clinical activity and presenting context-sensitive information within the environment, these systems could help teams maintain alignment as patient conditions, responsibilities, and procedural demands change.
Their technical capabilities, however, do not necessarily improve coordination.
TACT suggests that their value depends on whether computational representations remain consistent with clinical activity, role-specific displays preserve shared reference points, and clinicians retain the authority to question, correct, or override system interpretations.

Recent advances in AI may make some participant-envisioned functions more technically feasible, including summarizing prior activity, retrieving contextual guidance, and translating spoken updates into visual status cues.
However, because our study did not explicitly examine AI applications on AR-HMDs, we treat AI as one possible implementation mechanism rather than as a finding; in some cases, such as rapid onboarding, a structured and time-stamped account of prior activity may be sufficient without generated summaries.
Systems using automated inference would still need to make the source and uncertainty of information visible, preserve shared reference points, and allow clinicians to correct or override their outputs.

\subsection{Limitations and Future Work}
This study has several limitations.
First, the sample size of our study is low due to physiological and mental strain (i.e., burn out) experienced by healthcare workers in the global north; thus, our findings reflect the perspectives of a specific group of emergency care HCWs located in a well-developed country.
Our study, therefore,  may not capture practices across other institutions, specialties, countries, or staffing models.
Second, some participants had limited familiarity with AR-HMD technologies, relied on interviews and design scenarios rather than direct observation of clinical practice, which may have shaped how they imagined possible AR-HMD interfaces.
Nevertheless, our speculative design process accounted for this through video demonstrations and discussions about the capabilities and limitations of AR-HMDs. Accordingly, these findings identify perceived opportunities and constraints rather than demonstrating the feasibility, adoption, or effectiveness of a working AR-HMD system.

Although this study focuses on emergency medicine, information alignment may also help researchers examine coordination in other team-based domains.
For example, surgical teams, aviation crews, emergency response teams, and other time-critical groups also coordinate across partial and role-specific information.
Nevertheless, the mechanisms of alignment will likely differ across domains because teams use different communication practices, technologies, and authority structures.
Future work should examine how information alignment operates in these other settings before generalizing design recommendations beyond emergency care.

\section{Conclusion}
This work contributes an empirically grounded account of coordination breakdowns in action teams and introduces the Team Alignment and Coordination Taxonomy (TACT), which organizes critical alignment dimensions that lead to team failures, including information, expertise, procedural progress, and clinical judgment.
Across interviews, HCWs identified opportunities for AR-HMDs to facilitate pre-arrival briefing, procedural progress tracking, role-specific action, and verification of time-sensitive information, while emphasizing that such systems must preserve attention, hearing, movement, communication, and clinical authority.
TACT identifies where alignment must be maintained, how breakdowns can propagate across dimensions, and which relationships technological interventions should preserve.
 It therefore provides a foundation for designing ubiquitous systems that enhance high-stakes action teams without introducing additional burden or displacing human authority.
More broadly, the taxonomy can inform collaborative technology design across emergency medicine, aviation, construction, and disaster response.

\bibliographystyle{ACM-Reference-Format}
\bibliography{references}

\end{document}